\documentclass[twocolumn,showpacs,preprintnumbers,amsmath,amssymb,prb]{revtex4}

\usepackage{graphicx}
\usepackage{dcolumn}
\usepackage{bm}

\begin{document}

\title{Temperature dependence of coherent oscillations in
Josephson phase qubits}

\author{J. Lisenfeld$^1$, A. Lukashenko$^1$, M. Ansmann$^2$,
J. M. Martinis$^2$, and A. V. Ustinov$^1$}
\email{ustinov@physik.uni-erlangen.de} \affiliation{$^1$
Physikalisches Institut III, Universit{\"a}t
Erlangen-N{\"u}rnberg, D-91058 Erlangen, Germany\\
$^2$ Department of Physics and California Nanosystems Institute,
University of California, Santa Barbara, California 93106, USA }

\date{\today}

\begin{abstract}
We experimentally investigate the temperature dependence of Rabi
oscillations and Ramsey fringes in superconducting phase qubits
driven by microwave pulses. In a wide range of temperatures, we
find that both the decay time and the amplitude of these coherent
oscillations remain nearly unaffected by thermal fluctuations. The
oscillations are observed well above the crossover temperature
from thermally activated escape to quantum tunneling for undriven
qubits. In the two-level limit, coherent qubit response rapidly
vanishes as soon as the energy of thermal fluctuations $k_{\rm
B}T$ becomes larger than the energy level spacing $\hbar\omega$ of
the qubit. Our observations shed new light on the origin of
decoherence in superconducting qubits. The experimental data
suggest that, without degrading already achieved coherence times,
phase qubits can be operated at temperatures much higher than
those reported till now.
\end{abstract}

\maketitle

Superconducting qubits are electrical circuits based on Josephson
tunnel junctions which are fabricated using techniques borrowed
from conventional microelectronic circuits. Between various types
of rapidly developing superconducting qubits
\cite{Makhlin99,Devor-Wallr-Mart-04,Esteve-Vion-05,Wendin-Shumeiko-05},
advantages of Josephson phase qubits are their immunity to charge
noise in the substrate and simpler fabrication procedures due to
relatively large junction sizes. The experimentally controlled
degree of freedom in these devices is the superconducting phase
difference across a Josephson junction. By now, several groups
succeeded in demonstrating coherent oscillations and quantum state
manipulation for phase qubits
\cite{Martinis02,Claudon04,Wellstood,Lisenfeld-NTT-06}. A
significant increase of the coherence time in phase qubits became
possible after systematic research on microscopic defects in
insulator materials \cite{Martinis04a,Martinis04b,Martinis05b}.
These improvements led to the successful demonstration of quantum
state tomography tools for phase qubits
\cite{Martinis06b,Martinis06c}.

The decoherence effects in superconducting qubits give rise to a
decay of coherent oscillations in the population of the qubit
quantum states. One obvious reason for decoherence in Josephson
phase qubits are thermal fluctuations, whose effect has not yet
been studied. Most experiments so far have been performed at a
base temperature of a dilution refrigerator, typically around
15-30 mK, aiming at the lowest achievable temperature in order to
obtain the longest possible coherence times. Higher temperatures
remain thus unexplored. It is a common belief that
quantum-coherent behavior is limited to temperatures below the
crossover temperature $T^*$ between quantum tunneling and thermal
activation in a current-biased junction, typically of about 100
mK. In this paper we demonstrate that, in contrast the above
expectation, macroscopic quantum coherence can be observed at
temperatures much higher than $T^*$ of the junction.

We report here systematic measurements of the temperature
dependence of Rabi oscillations and Ramsey fringes in Josephson
phase qubits. We present experimental data for samples of
different origin and made of different materials. In a wide range
of temperatures, we find that both the decay time and the
amplitude of coherent oscillations are rather weakly affected by
temperature changes.

\begin{figure}
\includegraphics[width =8.5cm]{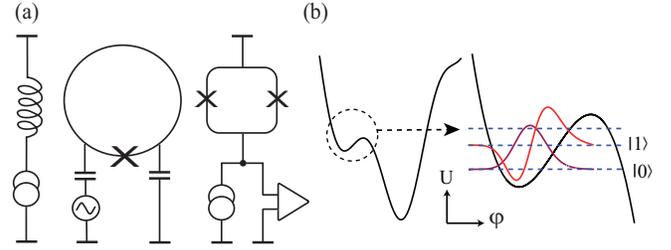} 
\caption{(a) Schematic of the phase qubit circuit. The qubit
junction is embedded in a superconducting loop which is coupled
inductively to the flux-biasing coil and readout dc-SQUID. (b)
Sketch of the qubit potential $U(\varphi)$. On the right side, a
zoom into the shallow left potential well indicates the wave
functions describing the two qubit states $|{0}\rangle$ and
$|{1}\rangle$. } \label{fig:schematic}
\end{figure}

\begin{figure*}
\includegraphics[width =18.0cm]{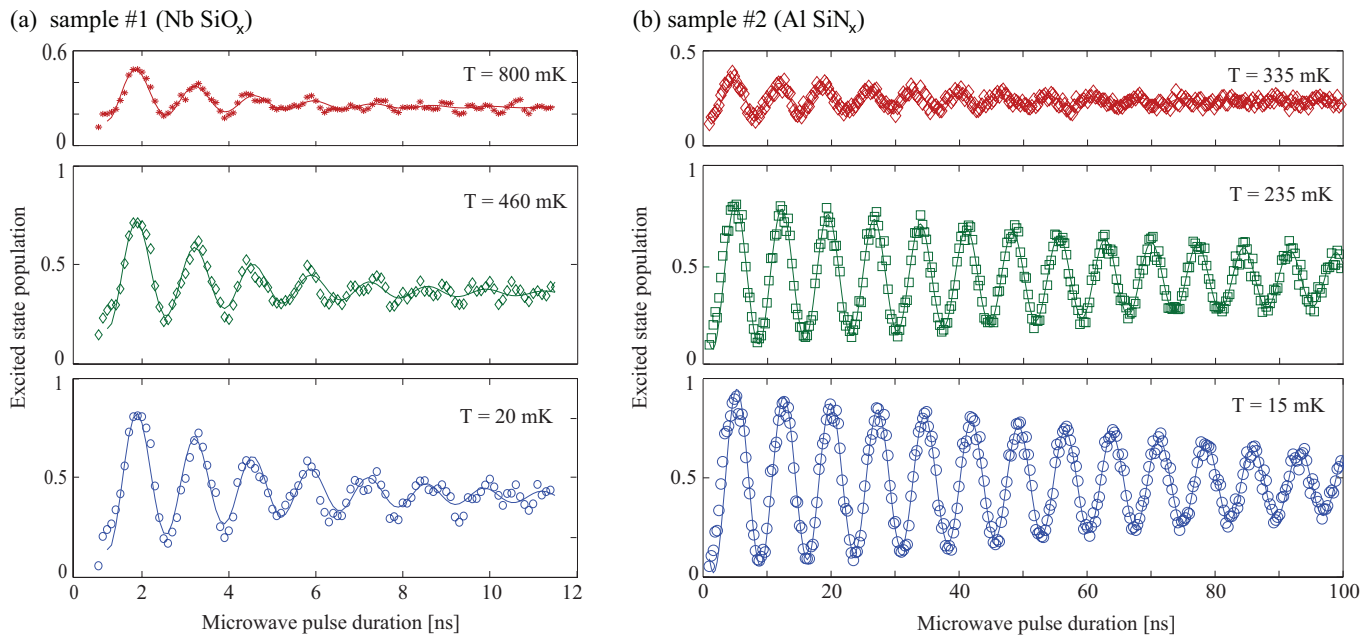} 
\caption{ (a) Rabi oscillations observed in Nb-based sample \#1
with SiO$_2$ insulation and (b) in Al-based sample \#2 featuring
SiN$_x$ insulation, at the indicated temperatures. Solid lines are
a fits to exponentially decaying sine functions from which Rabi
amplitude and decay time are extracted.} \label{fig:oscis}
\end{figure*}

A phase qubit uses as its logical quantum states $|{0}\rangle$ and
$|{1}\rangle$ the lowest two energy eigenstates in a metastable
potential well of the Josephson phase in a current-biased
junction. An elegant way to decouple the junction from its
electromagnetic environment is to apply the bias current through a
dc-transformer \cite{Martinis04a} by embedding the junction in a
superconducting loop as shown in Fig.~\ref{fig:schematic}(a). The
resulting circuit is known as rf-SQUID and has the potential
energy
\begin{equation}
\label{eq:pot} U(\varphi)=\frac{\hbar I_C}{2e}\left[ 1-\cos
\varphi + \frac{1}{2 \beta_L}
\left(\varphi-2\pi\frac{\Phi_\mathrm{ext}}{\Phi_0}\right)^2\right]
,\end{equation}
 where $\varphi$ is the phase difference across the junction,
$I_C$ is its critical current, $\Phi_\mathrm{ext}$ is the
externally applied flux through the qubit loop, and $\Phi_0$ is
the superconducting flux quantum.  The qubit loop inductance $L$
is chosen such that the parameter $\beta _L = 2\pi L I_C/ \Phi_0
\approx 4$, resulting in a double well potential plotted in
Fig.~\ref{fig:schematic}(b). By adjusting the external magnetic
flux close to $\Phi_0$, one of the wells can be made shallow
enough to contain only a small number of energy levels. The first
excited state $|{1}\rangle$ in this well can be populated by
resonant absorption of photons from the applied microwave field.
Reading out the state $|{1}\rangle$ population is done by using an
adiabatic dc-pulse of magnetic flux, which reduces the height of
the barrier separating the wells. The pulse amplitude is adjusted
such that an inter-well transition occurs only from the excited
state. Since the wells differ by the circulation direction of the
loop current, the transition from the shallow well to deep well
results in a change of magnetic flux, which is registered by
recording the switching current of an inductively coupled
dc-SQUID.

In our measurements, magnetic flux bias was generated by an
on-chip coil coupled weakly to the qubit inductance, while
microwave currents were supplied through a coplanar transmission
line connected capacitively to the qubit junction. The sample
temperature was stabilized in the range between 15 mK and 800 mK.
Magnetic shielding was provided by placing the sample in an
aluminum-coated copper cell surrounded by a superconducting lead
shield and two layers of $\mu$-metal. In order to reduce
electromagnetic interference, bias and microwave lines were
equipped with cold attenuators, while noise reduction in the
dc-SQUID wiring was achieved through capacitively shunted
copper-powder filters and current dividers at the 1K pot.

Samples of type \#1 have been fabricated according to our design
at two different foundries \cite{Hypres,VTT} by using standard
lithographic Nb/AlO$\mathrm{_x}$/Nb-trilayer processes. A
Josephson junction of area 7 ${\rm \mu m}^2$ with a current
density of 30 A/cm$^2$ was embedded into a two-turn loop of
inductance $L$=640 pH to form the qubit. All samples of type \#1
featured SiO${_2}$ as the insulation material between
superconducting layers surrounding the junctions.
Figure~\ref{fig:oscis}(a) shows Rabi oscillations of the excited
state population observed by varying the duration of an applied
resonant microwave pulse at $16.5$ GHz frequency followed by the
dc-readout pulse. Fitting the data by an exponentially damped sine
function, we extracted Rabi frequency, amplitude and decay time.
The $T_1$ time was obtained by measuring the decay rate of the
excited state population after a resonant $\pi$-pulse. All
measured type \#1 phase qubits showed rather short decoherence
times. As can be seen in Fig.~\ref{fig:oscis}(a), Rabi
oscillations have 70\% to 80\% visibility, decaying at a rate of
typically 3 to 5 ns.

In order to observe Rabi oscillation at higher temperatures it is essential
to avoid thermal activation out of the shallow potential well. The
activation rate increases as the potential barrier height becomes
comparable to the thermal energy $k_{\rm B} T$. By reducing the
field bias and operating in a deeper potential well, we observed
Rabi oscillations in the Nb-based type \#1 qubits at temperatures
up to 0.9 K. At a temperature of 0.8 K, the amplitude was reduced
by a factor of one half, while the oscillation decay time dropped
only to 80\%. In order to preserve high contrast of the readout at
high temperatures we reduced the amplitude of the readout pulse.
At high temperature, escape from the excited qubit state to the
deep well becomes possible not only by quantum tunnelling but also
by thermal activation over the barrier. Moreover, the width of the
measured switching-current histogram of the readout dc-SQUID
increases, which reduces the contrast between the two qubit
states. We separated the contributions of the two qubit flux
states at high temperatures by a fitting procedure using weighted
histograms of the two flux states.

Using the same experimental setup, we have also measured a sample
of type \#2 containing an Al/AlO$_x$/Al phase qubit fabricated as
described in Ref.~\onlinecite{Martinis06b}. This qubit features a
smaller junction of about 1 $\mu$m$^2$ area which is shunted by a
capacitor with SiN$_x$ dielectric layer. It was demonstrated in
Ref.~\onlinecite{Martinis05a} that replacing on-chip SiO${_2}$ by
SiN$_x$ reduces the number of parasitic two-level fluctuators and
significantly improves coherence time of the qubit. As it is seen
in Fig.~\ref{fig:oscis}(b), sample \#2 showed Rabi oscillations
decay times up to 100 ns -- more than one order of magnitude
longer than in Nb-SiO$_2$-based samples \#1. These data measured
in Erlangen are in good agreement with measurements at UC Santa
Barbara which were performed using samples of the same batch
\cite{Martinis06b}.

\begin{figure}
\includegraphics[width =8.5cm]{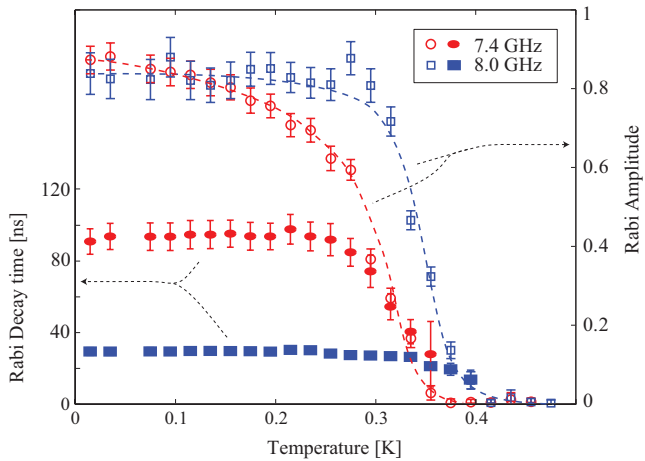}
\caption{Amplitude (right axis) and decay time (left axis) of Rabi
oscillation observed in sample \#2, plotted versus temperature.
Dashed lines are guides to the eye. } \label{fig:rabiamplitude}
\end{figure}

In Fig. \ref{fig:rabiamplitude}, we plot the temperature
dependence of the Rabi oscillation amplitude and decay time for
sample \#2. For each set of data points, we adjusted the magnetic
field bias in order to match the energy level spacing $\Delta E \equiv \hbar\omega$
to be at resonance with the chosen microwave frequency, which is indicated
in the legend. The microwave amplitude was chosen to result in Rabi
oscillation frequencies of $135$ MHz and $205$ MHz, respectively
for the two data sets at $7.4$ GHz and $8.0$ GHz driving
frequency. Remarkably, increasing the temperature up to about 300
mK, we observed nearly no effect on the oscillation amplitude and
decay time. At yet higher temperatures, the oscillations rapidly
decay and vanish completely at a temperature $T\approx T_\omega
\equiv \hbar \omega / k_{\rm B}$. The exact $T_\omega$ values are
$0.355$ K at $\omega = 2\pi\cdot 7.4$ GHz and $0.384$ K at
$\omega=2\pi\cdot 8.0$ GHz.

\begin{figure}[t]
\includegraphics[width =8.5cm]{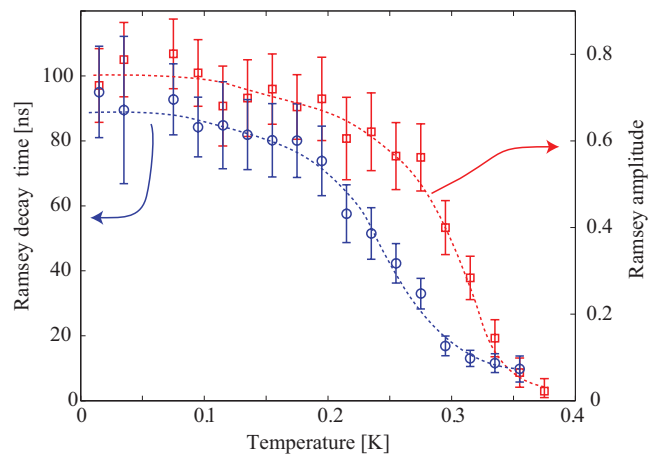}
\caption{Ramsey oscillation amplitude (right axis) and decay time
(left axis) vs. temperature (sample \# 2). Dashed lines are guides
to the eye.} \label{fig:ramseyamplitude}
\end{figure}

To measure the influence of temperature on the dephasing time
$T_2$ in sample \#2 we performed a Ramsey-type experiment, in
which two $\pi/2$-pulses separated by a variable duration are
applied to the qubit. As expected, the frequency of the observed
Ramsey fringes was equal to the detuning of the microwave from the
resonance frequency. We took a set of data at 335 MHz detuning
from the resonance at 8 GHz. Figure \ref{fig:ramseyamplitude}
shows the Ramsey oscillation amplitude and the extracted $T_2$
time versus temperature. At temperatures up to about 180 mK, we
measured a weak temperature dependence of the phase coherence
time, which at T = 15 mK was $T_2 \approx 90$ ns. At higher
temperatures, the time $T_2$ and fringe amplitude decrease and the
oscillations vanish at about 380 mK.

The theoretically expected temperature-induced decay rate from the
excited state to the ground state is \cite{Legget-RMP1996}
\begin{equation}
\label{eq:legget} \Gamma=\frac{2\pi\Delta
E}{\hbar}\frac{R_Q}{R}\,\left|A_{1,2}\right|^2 \left[1+{\rm
coth}\left(\frac{\Delta E}{k_{\rm B}T}\right)\right],
\end{equation}
where $R_Q=h/4e^2$ is the resistance quantum, $R$ is the effective
damping resistance, and
$A_{1,2}=\langle{0}|\frac{\phi}{2\pi}|{1}\rangle$ is the
interaction matrix element between the two states. The temperature
dependence of the decay rate is contained in the $\rm coth$-term
and is much less steep than the measured data shown in
Fig.~\ref{fig:rabiamplitude}. We can therefore qualitatively
interpret the measured temperature-independent decoherence in our
data as limited by microscopic two-level fluctuators at low
temperatures. Increasing temperature, on one hand, reduces the
fluctuators-induced decoherence as some part of the two-level
fluctuators gets saturated by energy absorbed from the thermal
bath. On the other hand, high enough temperature leads to
conventional decoherence of the qubit described by
Eq.~(\ref{eq:legget}), which damps coherent oscillations at
$T\approx T_\omega$. Detailed theoretical analysis of these
competing decoherence mechanisms goes beyond the scope of our
experimental work.

To directly measure the energy relaxation time $T_1$, we applied a
resonant microwave $\pi$-pulse that prepares the qubit in the
excited state. Then, after a variable delay time, we measure the
remaining population in the excited state. We observed exponential
decay at half-life times of about 4 ns for sample \#1 and about 90
ns for sample \#2. The life time $T_1$ is related to an
uncertainty in the energy of the excited state, giving rise to a
broadening $\delta E$ of the level at energy $E$. Consistent with
the measured $T_1$ relaxation times, we observed a spectroscopic
resonance width $\delta \omega$ of about $2\pi\cdot$400 MHz for
sample \#1 and about $2\pi\cdot$ 10 MHz for sample \#2. The
measured quality factors $Q= {\omega}/{\delta \omega} =
{E}/{\delta E}$ were about 30 at a transition frequency
$\omega_{01} = 2\pi\cdot 16.5$ GHz for the sample \#1 and about
600 at a frequency $\omega_{01} = 2\pi\cdot 8$ GHz for sample \#2.

The anharmonicity of the qubit potential has the consequence that
the separation between adjacent levels decreases with increasing
energy. An external microwave will hence be resonant with only one
transition if the individual levels are not too broad. This
requires a condition
\begin{equation}
\label{eq:levsepok} \hbar \omega_{12} + \frac{1}{2} (\delta E_1 +
\delta E_2) \ll \hbar \omega_{01} \end{equation} to be satisfied.
Here $\delta E_n$ is the full width of level $n$ arising from its
finite lifetime $\Gamma_n^{-1}$ and $\omega_{mn}$ is the
transition frequency from level $m$ to level $n$.

Due to their broad spectroscopic resonance width, our Nb-based
qubits of type \#1 can not be regarded as two-state systems. The
data shown in Fig.~\ref{fig:oscis}(a) thus correspond to
multilevel dynamics \cite{Claudon04,Wellstood}, which has been
recently shown to have its classical counterpart
\cite{NGJensen-Cirillo-PRL2005,Marchese-PRB2006}. In contrast, due
to the higher $Q$ value of the qubit \#2 (of about 600), in this
sample we couple by microwaves only the lowest two levels and thus
operate this qubit as a two-level system. Deeper potential well in
that case remains anharmonic enough not to result in poisoning of
the higher levels by microwaves applied at the frequency
$\omega_{01}$.

In conclusion, we presented measurements of multi-level Rabi
oscillations in low-$Q$ Josephson phase qubits fabricated using
standard niobium processes with SiO$_2$ dielectric. These
semi-classical Rabi oscillations persisted at temperatures up to
0.9 K and above, where only a modest decrease in decay time was
observed. In contrast, high-$Q$ aluminum-based phase qubits
featuring SiN$_x$ can be operated in the two-level limit even at
high temperature. Decoherence times measured in these quantum
systems depend very weakly on temperature up to the point
$T=T_\omega$, where the thermal energy $k_{\rm B}T$ becomes equal
to the energy level separation $\hbar \omega$. Our results
demonstrate that, without any significant degradation of their
coherence times, the best available phase qubits can be operated
at temperatures up to several 100 mK.

We acknowledge useful discussions with N. Gr{\o}nbech-Jensen, N.
Katz, J. E. Marchese, R.W. Simmonds and F. Wilhelm. Partial
financial support of this work by the Deutsche
Forschungsgemeinschaft (DFG) and European Aerospace Office of
Research and Development (EOARD) is acknowledged.


\begin{thebibliography}{0}

\bibitem{Makhlin99} Yu. Makhlin, G.~Sch{\"o}n, and A.~Shnirman,
Nature \textbf{398}, 305 (1999).

\bibitem{Devor-Wallr-Mart-04} M. H. Devoret, A. Wallraff, and
J. M. Martinis, cond-mat/0411174.

\bibitem{Esteve-Vion-05} D. Esteve and D. Vion, cond-mat/0505676.

\bibitem{Wendin-Shumeiko-05} G. Wendin and V.S. Shumeiko, cond-mat/0508729.

\bibitem{Martinis02} J. M. Martinis, S. Nam, J. Aumentado,
and C. Urbina, Phys. Rev. Lett. \textbf{89}, 117901 (2002).

\bibitem{Claudon04} J. Claudon, F. Balestro, F. W. J. Hekking, and
O. Buisson, Phys. Rev. Lett. \textbf{93}, 187003 (2004).

\bibitem{Wellstood} F. W. Strauch, S. K. Dutta, H. Paik, T. A. Palomaki,
K. Mitra, B. K. Cooper, R. M. Lewis, J. R. Anderson, A. J. Dragt,
C. J. Lobb, and F. C. Wellstood, cond-mat/0703081.

\bibitem{Lisenfeld-NTT-06} J. Lisenfeld, A. Lukashenko, and
A. V. Ustinov, unpublished (2006).

\bibitem{Martinis04a} R. W. Simmonds, K. M. Lang, D. A. Hite,
D. P. Pappas, and J.M. Martinis, Phys. Rev. Lett. {\bf 93}, 077003
(2004).

\bibitem{Martinis04b} K. B. Cooper, M. Steffen, R. McDermott, R. W.
Simmonds, S. Oh, D. A. Hite, D. P. Pappas, and J.M. Martinis,
Phys. Rev. Lett. {\bf 93}, 180401 (2004).

\bibitem{Martinis05b} J. M. Martinis, K. B. Cooper, R. McDermott, M. Steffen,
M. Ansmann, K. Osborn, K. Cicak, S. Oh, D.P. Pappas. R.W. Simmonds
and C.C. Yu, Phys. Rev. Lett. {\bf 95}, 210503 (2005).

\bibitem{Martinis05a} R. McDermott, R. W. Simmonds, M. Steffen,
K. B. Cooper, K. Cicak, K. Osborn, S. Oh, D. P. Pappas, and J. M.
Martinis, Science \textbf{307}, 5713 (2005).

\bibitem{Martinis06b} M. Steffen, M. Ansmann, R. McDermott,
N. Katz, R. C. Bialczak, E. Lucero, M. Neeley, E. M. Weig, A. N.
Cleland, and J. M. Martinis, Phys. Rev. Lett. \textbf{97}, 050502
(2006).

\bibitem{Martinis06c} M. Steffen, M. Ansmann, R. C. Bialczak,
N. Katz, E. Lucero, R. McDermott, M. Neeley, E. M. Weig, A. N.
Cleland, and J. M. Martinis, Science \textbf{313}, 1423 (2006).

\bibitem{Hypres} Hypres Inc., Elmsford, N.Y., USA.

\bibitem{VTT} VTT Technical Research Center, Finland.

\bibitem{Legget-RMP1996} A. J. Legget, S. Chakravarty, A.T. Dorsey
et. al., Rev. Mod. Phys. \textbf{59}, 1 (1996).

\bibitem{NGJensen-Cirillo-PRL2005} N. Gr{\o}nbech-Jensen and M. Cirillo,
Phys. Rev. Lett. \textbf{95}, 067001 (2005).

\bibitem{Marchese-PRB2006} J. E. Marchese, M. Cirillo, and
N. Gr{\o}nbech-Jensen, Phys. Rev. B \textbf{73}, 174507 (2006).

\end{thebibliography}
\end{document}